\documentclass[apj]{emulateapj}
\usepackage{natbib}
\usepackage{amsmath}
\usepackage{apjfonts}

\newcommand{\cm}{{\mathrm{cm}}}

\newcommand{\erg}{{\mathrm{erg}}}

\newcommand{\ergcc}{\erg\,\cm^{-3}}
\newcommand{\gcc}{\mathrm{g}\,\cm^{-3}}

\begin{document}

\title{The Dynamics and Afterglow Radiation of Gamma-Ray
  Bursts. I. Constant Density Medium}

\author{Weiqun Zhang and Andrew MacFadyen}
\affil{
  Center for Cosmology and Particle Physics, Physics
  Department, New York University, New York, NY 10003}

\begin{abstract}
  Direct multi-dimensional numerical simulation is the most reliable
  approach for calculating the fluid dynamics and observational
  signatures of relativistic jets in gamma-ray bursts (GRBs).  We
  present a two-dimensional relativistic hydrodynamic simulation of a
  GRB outflow during the afterglow phase, which uses the fifth-order
  weighted essentially non-oscillatory scheme and adaptive mesh
  refinement.  Initially, the jet has a Lorentz factor of 20.  We have
  followed its evolution up to 150 years.  Using the hydrodynamic
  data, we calculate synchrotron radiation based upon standard
  afterglow models and compare our results with previous analytic
  work.  We find that the sideways expansion of a relativistic GRB jet
  is a very slow process and previous analytic works have
  overestimated its rate.  In our computed lightcurves, a very sharp
  jet break is seen and the post-break lightcurves are steeper than
  analytic predictions.  We find that the jet break in GRB afterglow
  lightcurves is mainly caused by the {\emph{missing}} flux when the
  edge of the jet is observed.  The outflow becomes nonrelativistic at
  the end of the Blandford-McKee phase.  But it is still highly
  nonspherical, and it takes a rather long time for it to become a
  spherical Sedov-von Neumann-Taylor blast wave.  We find that the
  late-time afterglows become increasingly flatter over time.  But we
  disagree with the common notion that there is a sudden flattening in
  lightcurves due to the transition into the Sedov-von Neumann-Taylor
  solution.  We have also found that there is a bump in lightcurves at
  very late times ($\sim 1000$ days) due to radiation from the
  counter jet.  We speculate that such a counter jet bump might have
  already been observed in GRB 980703.
\end{abstract}


\section{Introduction}
\label{sec:intro}

In the standard fireball shock model for gamma-ray bursts (GRBs),
afterglows are due to synchrotron emission produced during the
slowdown of GRB outflows by surrounding media \citep[see e.g.,][for
recent reviews]{Zhang_Meszaros_2004_IJMPA, Piran_2005_RvMP,
  Meszaros_2006_RPPh, Granot_2007_RMxAC}.  The outflows of GRBs are
believed to be ultrarelativistic jets.  The deceleration of the
ultrarelativistic jet-like outflow in the early afterglow stage can be
well described by the spherical Blandford-McKee self-similar solution
\citep{Blandford_McKee_1976_PhFl}.  This is because the relativistic
beaming effect makes the material in the jet behave like an angular
patch of a spherical blast wave.  As it sweeps up the surrounding
medium, the jet decelerates and becomes less and less relativistic.
Hence the jet-like outflow will eventually expand sideways and become
increasingly spherical.  Thus the GRB outflow in the very late stages
can be described by the Sedov-von Neumann-Taylor non-relativistic
self-similar solution.  Unfortunately, there is no good analytic
solution that can describe the dynamics of the sideways expansion and
the transition from the ultrarelativistic phase to nonrelativistic
phase. 

The evolution of GRB outflows is an extremely important problem.  The
modeling of observable afterglow emission depends upon the dynamics of
the outflow.  An achromatic break in afterglow lightcurves is observed
in some GRBs \citep[e.g., GRB 990510,][]{Harrison_etal_1999_ApJ,
  Stanek_etal_1999_ApJ}.  The jet break is an indication that the GRB
outflow is jet-like \citep*{Rhoads_1999_ApJ, Sari_PH_1999_ApJ}.  To
understand the jet break, we must understand the multi-dimensional
dynamics of a relativistic jet.  The late-time radio afterglow can be
useful for an accurate estimate of the total energy of the GRB outflow
\citep{Frail_WK_2000_ApJ, Berger_KF_2004_ApJ, Frail_etal_2005_ApJ}.
Thus, the understanding of the transition from an ultrarelativistic
jet into a Newtonian spherical blast wave is crucial for correctly
interpreting the late-time radio observations.

The dynamics of GRB outflows is a multi-dimensional problem except
during the early Blandford-McKee and late Sedov-von Neumann-Taylor
phases.  Analytic approaches have been attempted to model the sideways
expansion based upon the simplified model of the jet as having a
top-hat structure as a function of angle during the expansion
\citep{Rhoads_1999_ApJ, Sari_PH_1999_ApJ,
  Panaitescu_Meszaros_1999_ApJ}.  In these models, the jet experiences
significant sideways expansion.  To include multidimensional effects
in an inexpensive way, \citet{Kumar_Granot_2003_ApJ} reduced the
multi-dimensional hydrodynamic equations to one dimension by
integrating over the radial profile of the jet.  They found little
sideways expansion of the jet.  However, the validity of these
approaches should be tested against full multi-dimensional
hydrodynamic simulations, which is certainly the most reliable
approach.  Unfortunately high-resolution multidimensional relativistic
hydrodynamic simulations are very expensive.  Thus far, very few
multidimensional hydrodynamic simulations have been performed to
investigate the dynamics of GRB outflows.
\citet{Granot_etal_2001_grba_conf} in their pioneering numerical work
found that GRB jets experience very little sideways expansion.  But
their simulation was not long enough to cover the transition into the
Sedov-von Neumann-Taylor solution.  A recent work by
\citet{Cannizzo_GV_2004_ApJ} has also found very little expansion in a
different initial setup, but their simulation suffered from severely
low resolution.

In this paper, we present a high-resolution relativistic hydrodynamic
simulation of the evolution of a GRB outflow during the afterglow
phase.  The evolution spans from when the Lorentz factor of the jet is
20 until 150 years after the burst.  This high-resolution simulation
is possible because our code, RAM \citep{Zhang_MacFadyen_2006_ApJS},
is massively parallel, utilizes adaptive mesh refinement (AMR), and
uses a fifth-order method.  The initial setup and results of the
hydrodynamic simulation will be presented in
Section~\ref{sec:dynamics}.  Based upon the hydrodynamic data and
standard afterglow models \citep*{Sari_PN_1998_ApJ,
  Granot_PS_1999_ApJ}, we have calculated synchrotron emission from
the simulated outflow (\S~\ref{sec:ag}).  We conclude with further
discussions of our results (\S~\ref{sec:dis}).

\section{Dynamics of GRB Outflows}
\label{sec:dynamics}

\subsection{Numerical Method and Initial Setup for Hydrodynamic
  Simulation} 
\label{sec:init}

The special relativistic hydrodynamic simulation in this paper was
performed with the RAM code \citep{Zhang_MacFadyen_2006_ApJS}.  RAM
utilizes the AMR tools in the FLASH code version 2.3
\citep{Fryxell_etal_2000_ApJS}, which in turn is a modified version of
the PARAMESH AMR package \citep{MacNeice_etal_2000_CoPhC}.  RAM
includes several modules to solve the special relativistic
hydrodynamics equations.  In the simulations in this paper, the
fifth-order weighted essentially non-oscillatory (WENO) scheme was
used. This scheme has been shown to achieve fifth-order accuracy for
smooth flows \citep{Zhang_MacFadyen_2006_ApJS} and excellent treatment
of shocks and contact discontinuities with no tunable numerical
parameters.

It is common in hydrodynamic simulations to use a constant gamma-law
equation of state (EOS) given by
\begin{equation}
P = (\hat{\gamma} - 1) \rho \epsilon,
\end{equation}
where $P$ is pressure, $\rho$ is mass density, $\epsilon$ is specific
internal energy density and $\hat{\gamma}$ is the adiabatic index all
measured in the fluid rest frame.  The adiabatic index can be assumed
to be $4/3$ and $5/3$ for relativistically hot and cold gas,
respectively, but it is usually set as a constant for entire
simulations.  The problem we studied in this paper, however, involves
both hot and cold gas.  GRB outflows are relativistically hot at the
beginning of the afterglow stage and become Newtonian in terms of both
fluid speed and sound speed.  The effect of the adiabatic index on GRB
afterglows can be quite large.  For example, the density jump due to
Newtonian strong shock compression is a factor of $4$ and $7$ for
$\hat{\gamma} = 5/3$ and $4/3$, respectively.  Hence the constant
gamma-law equation of state should be avoided for accurate treatment
of afterglow dynamics.  The exact equation of state for ideal gas,
which works for arbitrary temperature, was given by
\citet{Synge_1957_book}.  However, it is very expensive to use in
numerical simulations because it involves modified Bessel functions of
the second and third kinds.  To accurately follow the evolution of GRB
outflows with a reasonable cost, we used the TM EOS proposed by
\citet*{Mignone_PB_2005_ApJS}.  This equation of state has the correct
asymptotic behavior to $\hat{\gamma} = 5/3$ and $4/3$ in the
nonrelativistic and relativistic temperature limits, differs with the
exact equation of state by less than $4\%$ and can be solved at
negligible cost.

Our initial model is derived from the Blandford-McKee solution
\citep{Blandford_McKee_1976_PhFl}.  Given the energy of the blast
wave, $E_{\mathrm{iso}}$ and the density of the cold surrounding
medium, $\rho_0$, the blast wave can be fully described by a set of
self-similar relations that give the Lorentz factor, pressure and
density in the shocked fluid.  In our simulation, the energy in the
Blandford-McKee solution is set to $E_{\mathrm{iso}} = 10^{53}\,\erg$.
Note that this energy is not the total energy in the outflow.
Instead, it is the compensated energy the outflow would have if it
were a spherical blast wave.  In the simulation in this paper, we
adopted a number density of $1 \mathrm{cm}^{-3}$ for the
surrounding medium, which is consistent with GRB afterglow modeling
\citep[e.g.,][]{Panaitescu_Kumar_2001_ApJ, Panaitescu_Kumar_2002_ApJ}.
The surrounding medium is assumed to mainly consist of protons, and it
is assumed to be cold, $P_0 \ll \rho_0$.  Numerically, we set the
pressure to $P_0 = 10^{-10} \rho_0$, where units in which the speed of
light $c = 1$ are used in the simulation and henceforth in this paper.
The hydrodynamic simulation is started at the moment when the Lorentz
factor of the fluid just behind the shock is $\gamma_0 = 20$.  We have
run a two-dimensional axisymmetric simulation, in which the half
opening angle of the GRB jet is set to be $\theta_0 = 0.2$ radians.
Thus the total energy in the twin jet is $E_j \simeq 2.0 \times
10^{51}\,\erg$.  According to observations, a total energy of a few
times $10^{51}\,\erg$ is typical \citep[e.g.,][]{Frail_etal_2001_ApJ}.
A half opening angle of $0.2$ radians is also reasonable because it is
within the range of half opening angle inferred from observations
\citep[e.g.,][]{Frail_etal_2001_ApJ, Panaitescu_Kumar_2001_ApJ,
  Panaitescu_Kumar_2002_ApJ, Zeh_KK_2006_ApJ}.  We chose a slightly
large opening angle because of numerical reasons.  It is extremely
expensive to start our simulations from very early times when the
Lorentz factor of the jet is larger, because most of the energy in the
Blandford-McKee solution is concentrated in an extremely thin layer
behind the shock with a width of $\sim R/\gamma^2$ in the lab frame,
where $R$ is the radius of the shock.  The ``lab'' frame is the frame
in which the central engine is at rest and is equivalent to the
observer frame ignoring cosmological factors.  In our simulation, we
need to resolve the thin structures behind the shock in order to
accurately capture the true dynamics.  Furthermore, we want the
opening angle and Lorentz factor of the jet to satisfy the relation,
$\gamma_0 \gg 1/\theta_0$, so that the Blandford-McKee solution is
still valid for the jet.  Our choices of $\gamma_0 = 20$ and $\theta_0
= 0.2$ are thus reasonable.

The initial radius of the blast wave at the beginning of the
simulation is $R_0 \simeq 3.8 \times 10^{17}\,\cm$.  The Sedov
length for an explosion with an energy of $E_{j} \simeq 2.0 \times
10^{51}\,\erg$ into a medium with a density of $\rho_0 = 1.67\times
10^{-24}\,\gcc$ is
\begin{equation}
  l_{\mathrm{SNT}} \equiv
  \left(\frac{E_{j}}{(4\pi/3)\rho_0c^2}\right)^{1/3} \simeq
  6.8 \times 10^{17}\,\cm. 
\end{equation}
To simulate the evolution of the blast wave up to the Newtonian
regime, a large numerical box is necessary.  The size of the numerical
box for our two-dimensional simulation is $R_{\mathrm{MAX}} = 1.1
\times 10^{19}\,\cm$, which is about $16$ Sedov lengths.  Most of the
energy in the blast wave is initially concentrated in a thin layer
behind the shock with a width of $\Delta \sim 10^{15}\,\cm$ in the lab
frame.  It is thus very challenging to simulate the blast wave over
such a large dynamic range in lengthscale even with AMR.  At the
beginning of the simulation, the thin shell behind the shock occupies
a relatively small volume.  Thus we can afford to use very high
resolution initially.  As the blast wave expands, the volume of the
thin shell, which contains most of the energy, becomes larger.  Thus
AMR needs to create increasingly more numerical cells in order to
maintain the same high resolution of the structure, making the
simulation prohibitively expensive.  In our simulation, an algorithm
for derefinement of the adaptive mesh suggested by
\citet{Granot_2007_RMxAC} was used to save computing time.  In the
FLASH/PARAMESH AMR package, there is a parameter which controls the
maximal level of refinement and therefore decides the size of the
finest grid, which is usually fixed.  We change this parameter over
time according to an algorithm which utilizes a nice property of the
blast wave that the thin shell which needs to be resolved during the
relativistic regime behaves as $\Delta \sim R/\gamma^2 \sim t^4$,
where $t$ is the time in the lab frame.  Our algorithm reduces the
maximal refinement over time so that roughly the same number of cells
are used in the $r$-direction to resolve the radially widening shell at
different times without suffering from low resolution.

A spherical grid ($r, \theta$) with $0 \leq r \leq 1.1 \times
10^{19}\,\cm$ and $0 \leq \theta \leq \pi/2$ is employed in our
two-dimensional simulation.  Initially 16 levels of refinement is
used, and the finest cell has a size of $\Delta r \simeq 5.6 \times
10^{13}\,\cm$ and $\Delta \theta \simeq 9.6 \times 10^{-5}$.  At this
resolution, there are $2086$ cells at $\theta$-direction inside the
jet, and $17$ cells inside a shell with a width of $\Delta_0 =
R_0/\gamma_0^2$ behind the shock front.  If a uniform grid were used,
the total number of cells would have been more than $3$ billion in
order to achieve the highest resolution provided by AMR.  With AMR,
less than $7$ million cells are needed.  In our derefinement
algorithm, the maximal level of refinement decreases in response to
the change of the shell width until a specified minimal numerical
resolution is reached.  The minimal resolution is chosen such that AMR
uses at least 11 levels.

\subsection{Results of Two-Dimensional Hydrodynamic Simulation}
\label{sec:hydro}

\begin{figure*}
\plotone{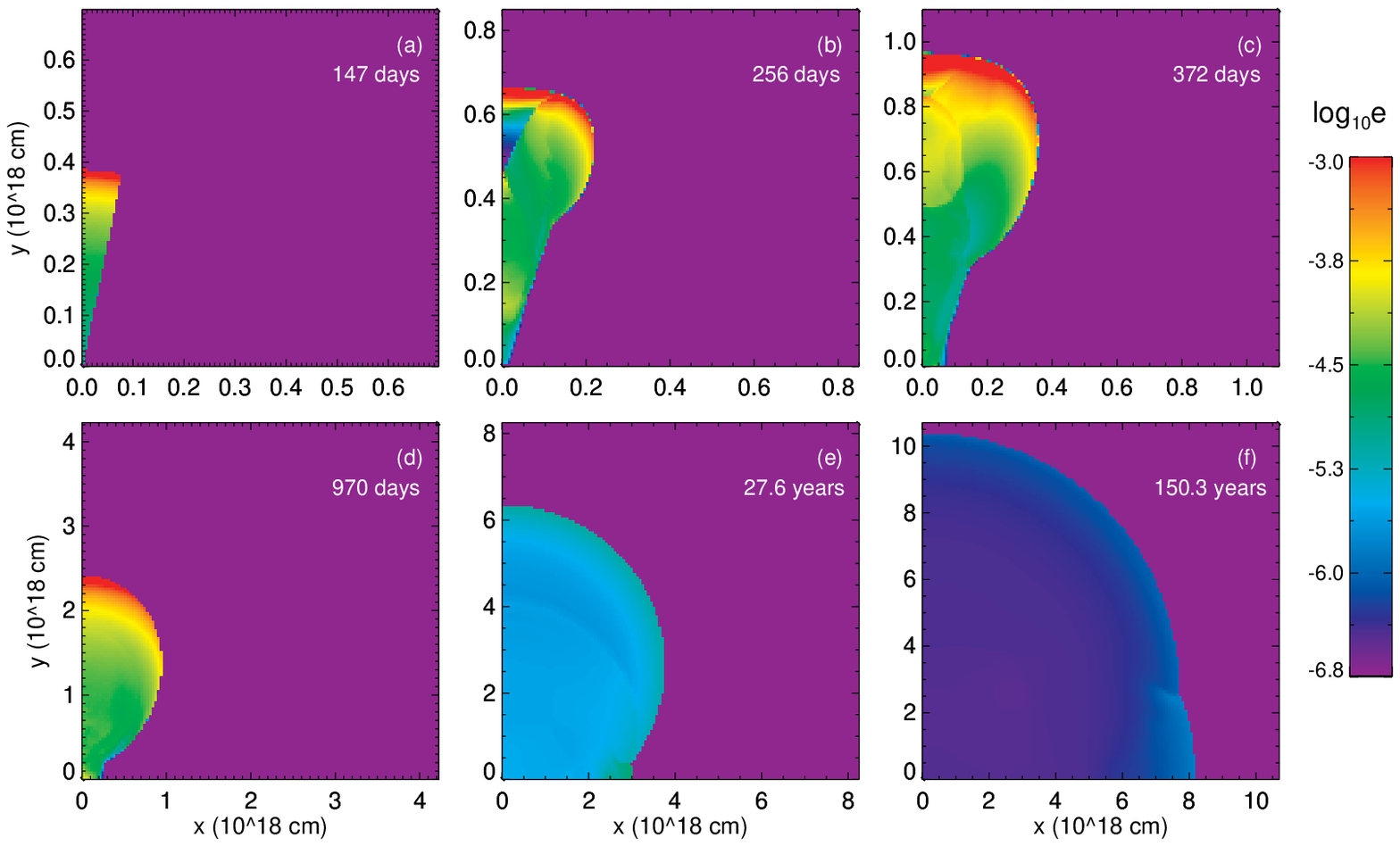}
\caption{Time evolution of the internal energy density in the local
  rest frame.  The internal energy density ($\ergcc$) is color coded
  on a logarithmic scale.  Snapshots of the simulation are shown at
  (a) 147 days, (b) 256 days, (c) 372 days, (d) 970 days, (e) 27.6
  years and (f) 150.3 years in the lab frame after the explosion.  The
  beginning of the simulation is at 147 days.  Panel~(d) shows that
  the GRB outflow is still highly 
  anisotropic at $t = t_{\mathrm{NR}} \approx 970$ days.  The
  nearly vertical shock front near the equator in 
  panels~(e) and (f) is a Mach stem, which is a result of the shock
  collision along the equator.   The minimal value in the color
  scale corresponds to $\leq 10^{-6.8}\,\ergcc$.   Note that the
  surrounding medium indeed has an 
  internal energy density of $\sim 2.25 \times 10^{-13}\,\ergcc$.
  \label{fig:sse}}
\end{figure*}

\begin{figure*}
\plotone{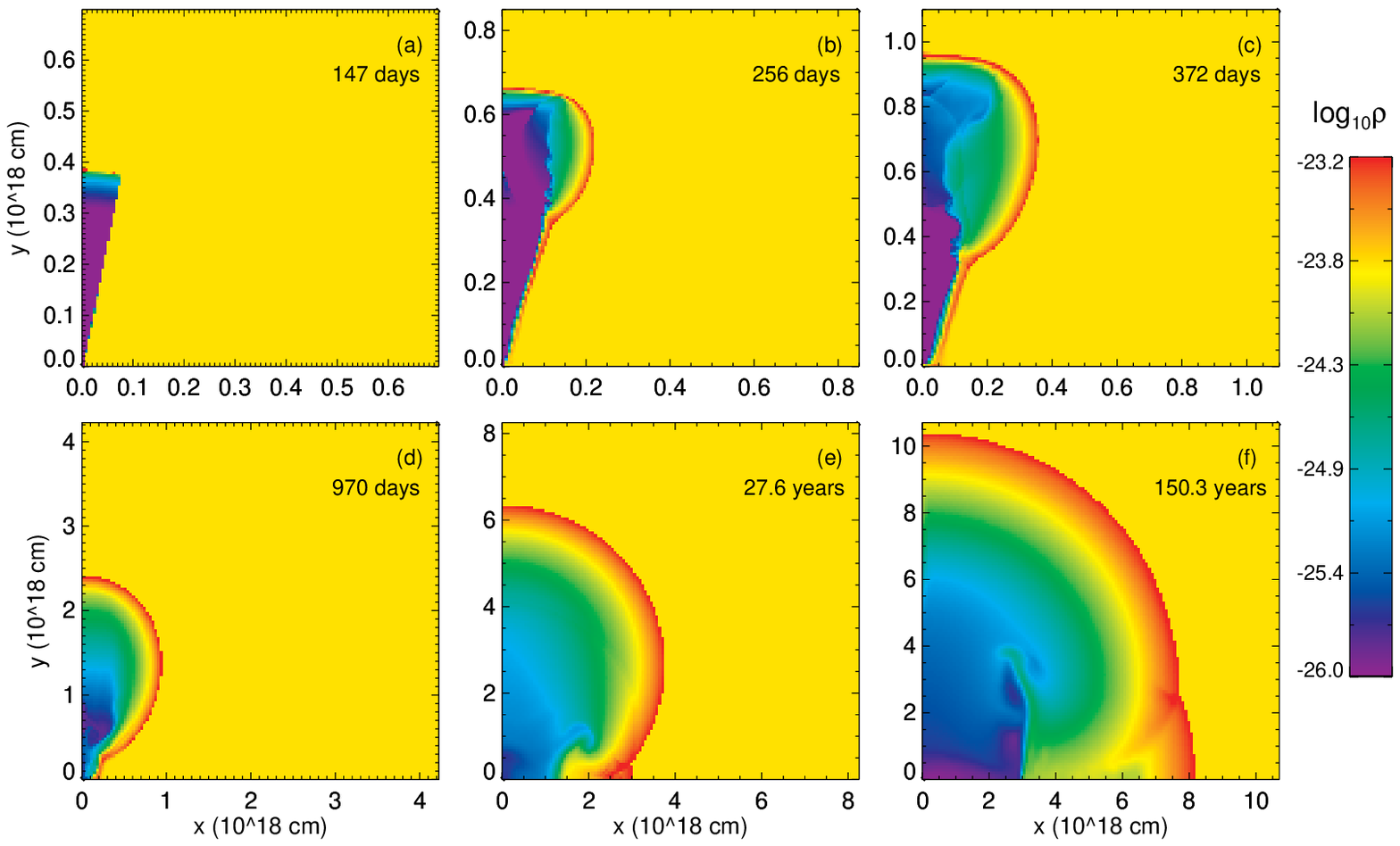}
\caption{Time evolution of the density in the local rest frame.  The
  density ($\gcc$) is color coded on a logarithmic scale.  Snapshots
  of the simulation are shown at (a) 147 days, (b) 256 days, (c) 372
  days, (d) 970 days, (e) 27.6 years and (f) 150.3 years in the lab
  frame after the explosion.  The beginning of the simulation is at
  147 days.  Panel~(d) shows that the GRB outflow is still highly
  anisotropic at $t = t_{\mathrm{NR}} \approx 970$ days.  The
  vortex feature in panels~(b) and (c) is an indication of the
  Kelvin-Helmholtz instability due to velocity shear.  The nearly
  vertical shock front near the equator in 
  panels~(e) and (f) is a Mach stem, which is a result of the shock
  collision along the equator. 
  \label{fig:ss}}
\end{figure*}

There are several time scales that are commonly used in the
literature.  These time scales are measured in the lab frame.  Note
that they are different from observer times. 
\begin{itemize}
\item $t_\theta$: This is the time at which the Lorentz factor behind
  the shock is equal to $\gamma = 1/\theta_0$ for a Blandford-McKee
  solution \citep{Blandford_McKee_1976_PhFl}.  It is given by
  \begin{equation}
    t_\theta \approx 373
    E_{\mathrm{iso},53}^{1/3} n_0^{-1/3}
    \left(\frac{\theta_0}{0.2}\right)^{2/3}\,\mathrm{days},\label{eq:ttheta}
  \end{equation}
  here $E_{\mathrm{iso},53}$ is the isotropic equivalent energy
  $E_{\mathrm{iso}}$ in units of $10^{53}\,\erg$ and $n_0$ is density
  of the medium in units of $\cm^{-3}$.
\item $t_s$: \citet{Livio_Waxman_2000_ApJ} argued that the transition
  from a jet to a spherical self-similar solution takes place over a
  time $ \Delta t_s \approx R_\theta /c$ where $R_\theta$ is the jet
  radius at time $t_{\theta}$.  Thus the outflow would become
  spherical at $t_s$, which is given by
  \begin{equation}
    t_s \approx t_\theta + R_\theta /c \approx 745
    E_{\mathrm{iso},53}^{1/3} n_0^{-1/3}
    \left(\frac{\theta_0}{0.2}\right)^{2/3}\,\mathrm{days}. \label{eq:ts}
  \end{equation}
\item $t_{\mathrm{SNT}}$: \citet{Livio_Waxman_2000_ApJ} defined the
  time $t_{\mathrm{SNT}}$ as the time at which the shock front moves
  at the speed of light assuming the blast wave is the Sedov-von
  Neumann-Taylor solution with an isotropic explosion energy of $E_j$.
  They argued that the GRB outflow becomes subrelativistic and can be
  described by the Sedov-von Neumann-Taylor solution after the time
  $t_{\mathrm{SNT}}$, which is given by
  \begin{equation}
    t_{\mathrm{SNT}} \approx 116     
    \left(\frac{E_{j}}{2 \times 10^{51}\,\erg}\right)^{1/3}
    n_0^{-1/3}\,\mathrm{days}. \label{eq:tsnt}
  \end{equation}
  Note that the time $t_{\mathrm{SNT}}$ is earlier that the initial
  time of our simulation $t_0$, when the Lorentz factor is $\gamma_0 =
  20$ for a relativistic jet obeying the Blandford-McKee solution.
  The jet has a Lorentz factor of 29 at $t_{\mathrm{SNT}}$.  Thus,
  the GRB outflow at $ t = t_{\mathrm{SNT}}$ is far from being
  described by the Sedov-von Neumann-Taylor solution.  The discrepancy
  is caused by assuming that the outflow is isotropic.
\item $t_{\mathrm{NR}}$: \citet{Piran_2005_RvMP} argued that the
  transition from relativistic to Newtonian should take place at
  $t_{\mathrm{NR}} = l_{\mathrm{NR}} / c$, where $l_{\mathrm{NR}} = (3
  E_{\mathrm{iso}} / 4\pi\rho_0c^2)^{1/3}$ is the Sedov length assuming
  that the GRB jet does not expand sideways.  After the transition the
  GRB outflow can be described by the Newtonian Sedov-von
  Neumann-Taylor solution.  The time is given by
  \begin{equation}
    t_{\mathrm{NR}} \approx 970
    E_{\mathrm{iso},53}^{1/3} n_0^{-1/3}\,\mathrm{days}. \label{eq:tnr}
  \end{equation}
  At the time $t_{\mathrm{NR}}$, the Lorentz factor behind the shock
  is $1.2$, if one assumes that the GRB does not expand sideways and
  the Blandford-McKee solution still applies.
\end{itemize}

It should be emphasized again that the above time scales are measured
in the lab frame.  They are different from observer times unless the
GRB outflow is already nonrelativistic.  We will further discuss these
time scales in Section~\ref{sec:frame}.

\begin{figure}
\plotone{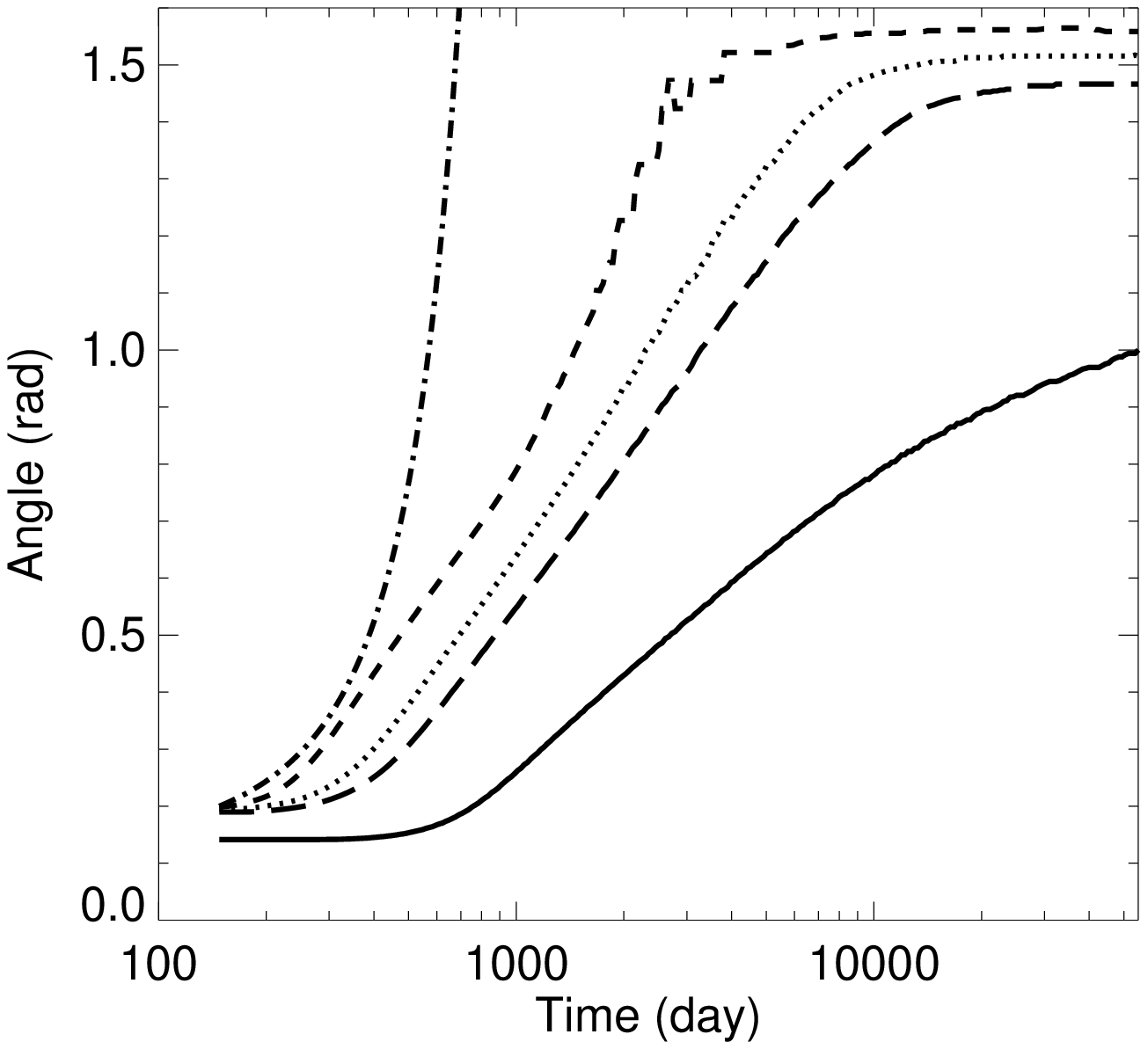}
\caption{Time evolution of the jet opening angle.  The opening angles
  inside which a certain percentage of the total energy, excluding
  rest mass energy, resides are shown for 50\% ({\it solid line}),
  90\% ({\it long dashed line}), 95\% ({\it dotted line}), and 99\%
  ({\it dashed line}) of the total energy.  Also shown in dash-dotted
  line is an analytic formula, $\theta = \theta_0
  e^{c(t-t_0)/l_{\mathrm{SNT}}}$.
  \label{fig:exx}}
\end{figure}

Figures~\ref{fig:sse} and \ref{fig:ss} show a series of snapshots of
the two-dimensional hydrodynamic simulation.  The simulation starts at
$t_0 \simeq 147$ days (measured in the lab frame of the burster) after
the initial explosion, and it was run up to $150$ years after the
explosion.  The surrounding medium has a density of $1.67 \times
10^{-24}\gcc$ and a specific internal energy of $2.25 \times
10^{-13}\,\ergcc$.  Initially the relativistic outflow propagates
mainly along the radial direction.  Later the jet will inevitably
undergo sideways expansion.  The structure of the flow is very
complicated.  In Panels~(b) and (c) of Figures~\ref{fig:sse} and
\ref{fig:ss}, we can identify a shock moving sideways, a rarefaction
wave propagating towards the jet axis, and a contact discontinuity in
between.  The contact discontinuity is Kelvin-Helmholtz unstable.
There is also a reverse shock inside the rarefaction wave propagating
towards the axis.  This is caused by the deceleration of the material
which moves sideways and sweeps up more and more surrounding medium.
It is shown in Panels~(b) and (c) of Figures~\ref{fig:sse} and
\ref{fig:ss} that the morphology of the GRB outflow at early times
consists of a jet cone which moves primarily in the radial direction
and a surrounding lobe which moves both radially and sideways.  At
late times, the GRB outflow becomes egg-like and then increasingly
spherical (Panels~(d), (e) and (f) of Figures~\ref{fig:sse} and
\ref{fig:ss}).  The nearly vertical Mach stem at the equator is due to
the collision of shocks from opposite hemispheres.  This feature is
unlikely to have direct observational consequences, but it is an
indication of the accuracy of the simulation.

What is important for observations is the distribution of the bulk of
the energy.  Since the jet material near the forward shock front
initially moves at nearly the speed of light, very little sideways
expansion takes place for the ultrarelativistic material due to
relativistic kinematics.  Inside the jet, the Lorentz factor decreases
radially inwards.  The mildly relativistic and Newtonian jet material
behind the forward shock front undergoes more sideways expansion as
shown in Figures~\ref{fig:sse} and \ref{fig:ss}.  Also shown in
Figures~\ref{fig:sse} and \ref{fig:ss} are snapshots of internal
energy density and mass density at $t = t_{\mathrm{NR}} \approx
970\,\mathrm{days}$.  It should be noted that the GRB outflow at this
moment is still highly anisotropic in the angular distribution of
energy.

Figure~\ref{fig:exx} shows the time evolution of the opening angles
inside which a certain percentage of the total energy, excluding rest
mass energy, resides.  At $t = t_{\theta} \approx 373$ days, 50\%,
90\%, 95\% and 99\% of the total energy are inside an opening angle of
0.14, 0.24, 0.28 and 0.41, respectively.  At $t = t_{\mathrm{NR}}
\approx 970$ days, 50\%, 90\%, 95\% and 99\% of the total energy is
within an opening angle of 0.25, 0.54, 0.63 and 0.77, respectively.
At $t = 20$ years, 50\%, 90\%, 95\% and 99\% of the total energy is
within an opening angle of 0.72, 1.28, 1.43 and 1.55, respectively.
Note that for a spherical blast wave, the opening angles would be 1.05,
1.47, 1.52 and 1.56, for 50\%, 90\%, 95\% and 99\% of the total
energy, respectively.  It is clear that the spreading of energy to
large angles happens very slowly.  It takes $\sim 10-20$ years for the
jet to be somewhat spherical.  The approach to a spherical blast
wave and the Sedov-von Neumann-Taylor solution is a very slow process.

\begin{figure}
\plotone{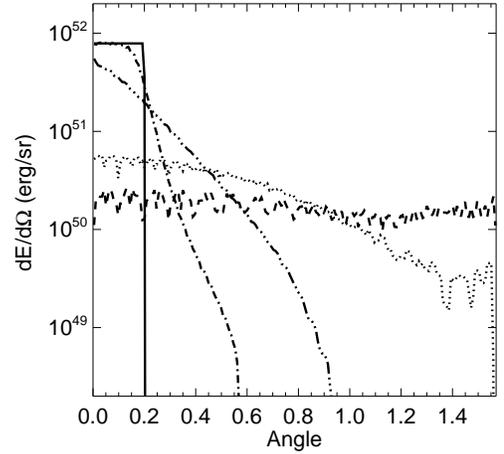}
\caption{Angular distribution of energy.  Different lines are for
  different times: $t = 0$ ({\it solid line}), $t_\theta \approx
  373$ days ({\it dash-dotted line}), $t_{\mathrm{NR}}
\approx 970$ days ({\it dash dot dot line}), $5 t_{\mathrm{NR}}
\approx 13$ years ({\it dotted line}) and $150$ years ({\it dashed
  line}).  
  \label{fig:dedo}}
\end{figure}

The angular distribution of the total energy, excluding rest mass
energy at $t = 0$, $t_\theta \approx 373$ days, $t_{\mathrm{NR}}
\approx 970$ days, $5 t_{\mathrm{NR}} \approx 13$ years and $150$
years is shown in Figure~\ref{fig:dedo}.  The outflow evolves from a
jet with an opening angle of $\theta_0 = 0.2$ into an isotropic blast
wave.  At $t = 5 t_{\mathrm{NR}} \approx 13$ years, the energy per
unit sold angle varies by about an order of magnitude between the
axis and the equator.  During the evolution, the angular energy
distribution is not in any kind of universal profile.  In particular,
the outflow does not resemble a top-hat jet.

\begin{figure}
\plotone{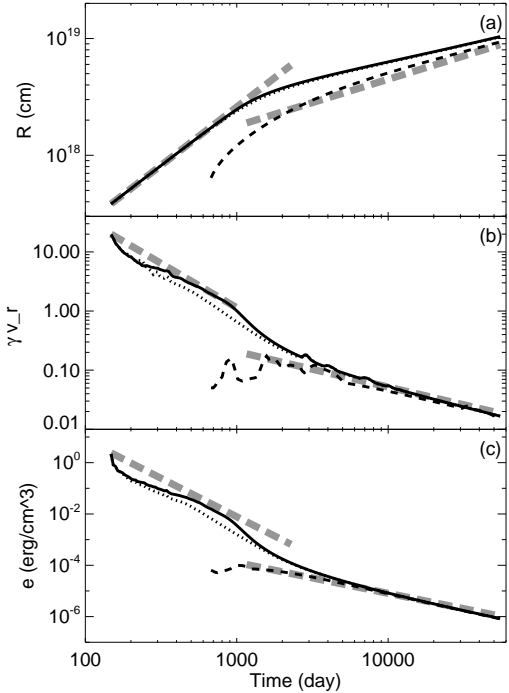}
\caption{Properties of the fluid just behind the forward shock as a
  function of time.  The 
  position of the forward shock, the product of the Lorentz factor and
  radial velocity, and internal energy density are plotted in Panels~(a)
  ({\it top}), (b) ({\it middle}), and (c) ({\it bottom}),
  respectively.  Properties at three different angles are shown: (1)
  $\theta = 0$ ({\it solid lines}), (2) $\theta = 0.19$ ({\it dotted
    lines}), and (3) $\theta = \pi/4$ ({\it dashed lines}).  In
  Panel~(b), the unit for velocity is the speed of light.  The
  Blandford-McKee solution and Sedov-von Neumann-Taylor solution
  are also shown in thick gray dashed lines.  In Panel~(b), instead of
  the product of the Lorentz factor and velocity, the Lorentz factor
  and velocity are plotted, for the Blandford-McKee solution and
  Sedov-von Neumann-Taylor solution, respectively. 
  \label{fig:slice}}
\end{figure}

The very little sideways expansion we find is in agreement with the
results of previous numerical simulations
\citep{Granot_etal_2001_grba_conf, Cannizzo_GV_2004_ApJ}.
Figure~\ref{fig:exx} shows that the opening angle grows
logarithmically over time.  However, analytic work
\citep*{Rhoads_1999_ApJ, Sari_PH_1999_ApJ} has predicted that the jet
opening angle should grow exponentially, $\theta \sim
e^{c(t-t_0)/l_{\mathrm{SNT}}}$ (Fig.~\ref{fig:exx}).  Thus, according
to these analytic estimates, the transition from jet-like to
spherical-like takes place over practically no time.  Why does the jet
spread so rapidly in the analytic work compared with that in the
numerical simulations?  The main reason is that the jet in analytic 
work is assumed to have an unrealistic top-hat distribution as a
function of angle.  As it expands sideways, the top-hat jet has more
and more working surface, which in turn rapidly decreases the Lorentz
factor.  Since the speed of the sideways expansion depends upon the
Lorentz factor, the sideways expansion of a top-hat jet becomes a
runaway process, which grows exponentially.  In fact, the jet is far
from top-hat during the sideways expansion (Fig.~\ref{fig:dedo}).  The
information of the existence of surrounding medium outside the jet
opening angle propagates towards the axis as a rarefaction wave, which
moves at the sound speed in the local rest frame.  The part of the jet
that the rarefaction has not reached will behave exactly like a
spherical outflow and continue to expand radially.  The part of the
jet that is affected laterally by the jet edge is inside a rarefaction
wave, which has a more complicated angular profile than a top-hat.
Moreover, the reverse shock which appears in the early stages further
slows down the sideways expansion.  Can numerical simulations
underestimate the rate of sideways expansion?  We believe that it is
unlikely.  In fact, numerical simulations tend to overestimate the
rate of sideways expansion of relativistically moving material due to
the inevitable numerical viscosity \citep{Zhang_MacFadyen_2006_ApJS}.
Also the Blandford-McKee solution is very challenging for numerical
simulations due to its extremely thin structure behind the shock.  The
Lorentz factor of the material just behind the shock tends to be lower
than the analytic Blandford-McKee solution (Fig.~\ref{fig:slice}).
This will also increase the rate of sideways expansion numerically.
Hence, we conclude that the runaway lateral expansion derived in the
analytic work does not exist in reality and the sideways expansion is
a slow process.

Figure~\ref{fig:slice} shows the time evolution of various properties
of the fluid just behind the forward shock: the position of the shock
front, the product of the Lorentz factor and radial velocity, and the
internal energy density at various angles: $\theta = 0$, $0.19$, and
$\pi/4$.  Also shown are the Blandford-McKee solution and Sedov-von
Neumann-Taylor solution for comparison.  It is shown that the outflow
can be approximately described by the Blandford-McKee solution at
early times ($t < t_{\mathrm{NR}} \approx 970$ days) and the Sedov-von
Neumann-Taylor solution at late times ($t > 5 t_{\mathrm{NR}} \approx
5000$ days).  And the transition takes place over a period of $\sim 4
t_{\mathrm{NR}}$.  It is striking that the Blandford-McKee solution is
valid for the material near the jet axis until the time when the
Lorentz factor decreases almost to 1.  Note that the assumption of
ultrarelativistic velocity upon which the Blandford-McKee solution is
based has become invalid before that time.  It should also be noted
that even with 16 levels of mesh refinement, the simulation still
suffers from insufficient resolution at early times.

The results of our hydrodynamic simulation are summarized as follows:
(1) The initial condition of the jet is described by the
Blandford-McKee solution; (2) The jet slows down as it sweeps up the
surrounding medium; (3) At $t \sim t_\theta$, the jet starts to
undergo sideways expansion, but at a rate much slower than the
exponential growth predicted by analytic work; (4) The jet becomes
nonrelativistic and the Blandford-McKee solution breaks down at $t
\sim t_{\mathrm{NR}}$; (5) The outflow becomes more and more
spherical and undergoes a {\emph{slow}} transition into the
Sedov-von Neumann-Taylor solution; (6) After $t \sim 5
t_{\mathrm{NR}}$, the outflow is close to spherical and can be
described by the Newtonian Sedov-von Neumann-Taylor solution.

\section{Afterglow Radiation of GRBs}
\label{sec:ag}

We now calculate the afterglow radiation of GRBs using the data from
our two-dimensional special relativistic hydrodynamic simulation.

\begin{figure}
\plotone{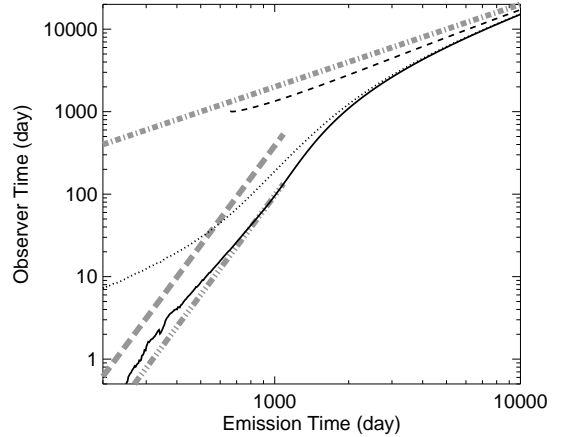}
\caption{Observer time vs. Emission time.  The results for fluid
  elements just behind the shock at various angles are shown: $\theta
  = 0$ ({\it solid line}), 0.19 ({\it dotted line}), and $\pi/4$ ({\it
    dashed line}).  The relations $t_\earth = (1+z) t/4\gamma^2$ ({\it
    long dashed gray line}) and $t_\earth = (1+z) t/16\gamma^2$ ({\it
    dash dot dot gray line}) are plotted for comparison.  Here,
  $\gamma$ is assumed to obey the Blandford-McKee solution, and the
  cosmological redshift is set to $z=1$.  Also plotted is $t_\earth = (1+z) t$
  ({\it dash-dotted gray line}).\label{fig:tt}}
\end{figure}

\subsection{Frames and Times} 
\label{sec:frame}

Two frames are involved in the hydrodynamic simulation
(\S~\ref{sec:hydro}): the local rest frame of a fluid element and the
lab frame of the GRB central engine.  The measurements of spacetime in
the above two frames satisfy the Lorentz transformation.  For an
observer on the earth, there is an additional frame: the observer
frame \footnote{The observer frame is the same inertial frame as the
  lab frame if cosmological effects are neglected
  \citep{Zhang_Meszaros_2004_IJMPA}.}.  If a photon is emitted at time
$t$ and position $\vec{r}$ in the lab frame, an observer will receive
it at
\begin{equation}
  t_\earth = (1+z) (t - \frac{\vec{r} \cdot \vec{n}}{c} ) \label{eq:tearth}
\end{equation}
after the main burst.  Here, $z$ is the redshift of the GRB, and
$\vec{n}$ is the unit vector pointing from the burster to the earth.
Clearly, photons received by the observer at a given time are emitted
by different regions of the GRB outflow at different times.  However,
the emission is mostly from a very small region if the outflow is
relativistic.  Therefore, a simple relation is often used in analytic
work for the emission time and observer time.  For a GRB outflow in
the relativistic phase, the two times satisfy
\begin{equation}
  t_\earth \approx (1+z) \frac{t}{4\gamma^2},
  \label{eq:ttr}
\end{equation}
here $\gamma$ is the Lorentz factor of the fluid just behind the
shock, which is described by the Blandford-McKee solution.  The factor
of 4 in the denominator of Equation~\ref{eq:ttr} is due to the effect
that ``typical'' photons seen by an observer are emitted from off-axis
regions \citep[e.g.,][]{Waxman_1997b_ApJ}.  When the velocity of the
outflow is Newtonian (i.e., $v \ll c$), the relation is simply
\begin{equation}
  t_\earth \approx (1+z) t.  \label{eq:ttn}
\end{equation}
There is, however, no good analytic
relation for the transrelativistic phase.

Our numerical approach has the advantage of treating the various times
accurately.  Figure~\ref{fig:tt} shows the observer time versus
emission time for fluid elements just behind the shock at various
angles.  Analytic relations are also plotted for comparison.  We have
discussed various timescales in \S~\ref{sec:hydro}, including
$t_\theta$, $t_s$, $t_{\mathrm{SNT}}$ and $t_{\mathrm{NR}}$
(Eqs.~\ref{eq:ttheta}, \ref{eq:ts}, \ref{eq:tsnt} and \ref{eq:tnr}).
All these timescales are measured in the lab frame.  A good
approximation is that Eq.~\ref{eq:ttr} is valid for $t_\theta$.  The
time $t_s$ consists of two parts (Eq.~\ref{eq:ts}).  Since the outflow
is relativistic before $t_\theta$ and is then assumed to become
Newtonian very quickly due to the sideways expansion,
\citet{Livio_Waxman_2000_ApJ} argued that the GRB outflow will become
spherical at $t_{s,\earth} \sim (1+z) R_\theta/c \sim 372 (1+z)
E_{\mathrm{iso},53}^{1/3} n_0^{-1/3}
(\theta_0/0.2)^{2/3}\,\mathrm{days}$.  The time
$t_{\mathrm{NR},\earth} \sim (1+z) t_{\mathrm{NR}} \sim 970 (1+z)
E_{\mathrm{iso},53}^{1/3} n_0^{-1/3}\,\mathrm{days}$ is also often
assumed to be the beginning of the Sedov-von Neumann-Taylor phase
\citep[e.g.,][]{Piran_2005_RvMP}.  But it is clearly shown in
Figure~\ref{fig:tt} that the relation $t_\earth = (1+z) t$ cannot be
used at times $t_s$ and $t_{\mathrm{NR}}$.  It would be a mistake to
treat them as the observer times (after a cosmological redshift
correction.)  Another mistake is that the outflow at these times is
neither spherical nor Sedov-von Neumann-Taylor as we have shown in
\S~\ref{sec:hydro}.  The transition to the spherical Sedov-von
Neumann-Taylor solution takes place over a rather long period ($\sim 5
t_{\mathrm{NR}}$) in the lab frame.  But, it turns out that the two
mistakes somewhat compensate each other.  At $t_\earth = (1+z)
t_{\mathrm{NR}}$ , the observed radiation is emitted by
nonrelativistic material, which is undergoing the transition into the
spherical Sedov-von Neumann-Taylor phase.  However, $(1+z)
t_{\mathrm{SNT}} \approx 116 (1+z) (E_{j}/2 \times
10^{51}\,\erg)^{1/3} n_0^{-1/3}\,\mathrm{days}$ is certainly
inappropriate as the observer time for the beginning of either the
nonrelativistic phase or the Sedov-von Neumann-Taylor phase.

\subsection{Standard Afterglow Model}
\label{sec:agmodel}

Our calculation of the radiation is based upon the standard external
shock model for GRB afterglows \citep{Meszaros_Rees_1997_ApJ,
  Wijers_RM_1997_MNRAS, Waxman_1997a_ApJ, Waxman_1997b_ApJ,
  Sari_PN_1998_ApJ}.  We use the formalism introduced by
\citet{Granot_PS_1999_ApJ}, which works nicely with data from
numerical simulations.  A large number of data dumps from the
hydrodynamic simulation are stored.  The numerical GRB outflow
consists of many small fluid elements $\Delta V = 2 \pi r^2
\sin{\theta} \Delta r \Delta \theta$ at a discrete set of lab times.
The velocity, number density and internal energy density of each fluid
element is known at each lab time.  A fluid element is assumed to
exist over a period of $\Delta t$ depending upon the interval between
dumps.  The observed flux density at $t_\earth$ given by
Eq.~\ref{eq:tearth} over a period of $\Delta t_\earth = (1+z) \Delta
t$ due to radiation emitted by the fluid element is given by,
\begin{equation}
  \Delta F(\nu_\earth, t_\earth) = \frac{1+z}{4\pi d_L^2} 
    \frac{P^{\prime}(\nu^{\prime})}{\gamma^2 (1 -
      \vec{\beta} \cdot \vec{n})^2} \Delta V,  \label{eq:f}
\end{equation}
here $d_L$ is the luminosity distance, $\gamma$ is the Lorentz factor,
$\vec{\beta}$ is the dimensionless velocity of the fluid element, and
$P^{\prime}(\nu^{\prime})$ is the emitted energy per unit volume per
unit frequency per unit time measured in the fluid rest frame, where
the frequency $\nu^{\prime}$ is also measured in the fluid rest frame.
The frequencies measured in the fluid rest frame and observed on the
earth are related by \footnote{There is a typo in Eq.~4 of
  \citet{Granot_PS_1999_ApJ}.  There should be a factor of $1+z$ in
  the first argument of $P^{\prime}$.}
\begin{equation}
  \nu^{\prime} = (1+z) \gamma (1 -
  \vec{\beta} \cdot \vec{n}) \nu_\earth. 
\end{equation}
The total observed flux density as a function of frequency and
observer time, $F(\nu_\earth, t_\earth)$, can be calculated by
combining $\Delta F(\nu_\earth, t_\earth)$ over all the fluid elements
at all discretized times.

We consider synchrotron emission, but ignore synchrotron
self-absorption and inverse Compton scattering, for the sake of
simplicity.  To compute synchrotron emission of a fluid element with a
number density $n$ and an internal energy density $e$, we use the
standard approach of assuming that $e_e = \epsilon_e e$ and $e_B =
\epsilon_B e$, here $e_e$ is the energy density of radiation emitting
electrons, $e_B$ is the energy density of the magnetic field, and
$\epsilon_e$ and $\epsilon_B$ are two parameters.  A power-law
distribution is assumed for the radiation emitting electrons:
$N(\gamma_e) \sim \gamma_e^{-p}$, where $\gamma_e$ is the Lorentz
factor of electrons, and $p = 2.5$ is a constant parameter.  We follow
the work of \citet{Sari_PN_1998_ApJ} for calculating synchrotron
emission.  For our purposes, it is sometimes more convenient to use the
lab frame time.  We have taken into account the effect of electron
cooling on the spectral power of synchrotron emission.  The critical
value of the Lorentz factor of electrons is computed using
\begin{equation}
  \gamma_c = \frac{3 m_e c \gamma}{4 \sigma_T e_B t}, \label{eq:gamc}
\end{equation}  
here $m_e$ is the mass of an electron, $\sigma_T$ is the Thompson
cross section, $\gamma$ is the Lorentz factor of the fluid element,
and $t$ is the time in the lab frame.  The spectral power of
synchrotron emission is assumed to be a broken power-law with three segments
separated by $\nu_c$, the cooling frequency, and $\nu_m$, the typical
frequency of electrons with the minimal Lorentz factor.  Assuming that
the outflow obeys the Blandford-McKee solution and the emission time
and observer time are related by Eq.~\ref{eq:ttr}, the two break
frequencies in the observer frame are given by\footnote{Our
  Eq.~\ref{eq:nuc} for $\nu_c$ is 16 times smaller than Eq.~11 in
  \citet{Sari_PN_1998_ApJ} due to a factor of 4 difference in the
  expression for $\gamma_c$ between our Eq.\ref{eq:gamc} and Eq.~6 in
  \citet{Sari_PN_1998_ApJ}.  }
\begin{eqnarray}
\nu_{c,\earth} &=& 1.8 \times 10^{11} (1+z)^{-1} \epsilon_B^{-3/2}
n_0^{-3/2} \left(\frac{t}{147\,\mathrm{days}}\right)^{-2}\,\mathrm{Hz} \nonumber\\
 &=& 5.3 \times 10^{10} (1+z)^{-1/2} \epsilon_B^{-3/2} E_{\mathrm{iso},53}^{-1/2}
 n_0^{-1} t_{\earth,d}^{-1/2}\,\mathrm{Hz} \label{eq:nuc}
\end{eqnarray}
and 
\begin{eqnarray}
\nu_{m,\earth} &=& 6.7 \times 10^{16} (1+z)^{-1} \epsilon_B^{1/2}
\epsilon_e^2 E_{\mathrm{iso},53}^2 n_0^{-3/2}
\left(\frac{t}{147\,\mathrm{days}}\right)^{-6}\,\mathrm{Hz} \nonumber\\
&=& 1.8 \times 10^{15} (1+z)^{1/2} \epsilon_B^{1/2}
\epsilon_e^2 E_{\mathrm{iso},53}^{1/2} t_{\earth,d}^{-3/2}\,\mathrm{Hz}, \label{eq:num}
\end{eqnarray}
here $t_{\earth,d}$ is the observer time in units of days.  The peak
flux density is given by
\begin{equation}
  F_{\nu,\mathrm{max}} = 970 (1+z) \epsilon_B^{1/2} E_{\mathrm{iso},53} n_0^{1/2}
  d_{L,28}^{-2}\,\mathrm{mJy}.  
\end{equation}
Here, a factor of 0.88 was added to the expression for
$F_{\nu,\mathrm{max}}$ \citep[Eq.~11 in][]{Sari_PN_1998_ApJ} to
reflect the fact that the synchrotron emission at the peak frequency
has a smoothly curved shape rather than a broken power-law shape
\citep{Granot_PS_1999_ApJ}.  The two break frequencies become equal to
the critical frequency
\begin{equation}
  \nu_{t,\earth} = 2.9 \times 10^{8} (1+z)^{-1} \epsilon_B^{-5/2}
  \epsilon_e^{-1} E_{\mathrm{iso},53}^{-1}n_0^{-3/2}\,\mathrm{Hz} 
\end{equation}
at
\begin{equation}
  t_t = 3.6 \times 10^{3} \epsilon_B^{1/2} \epsilon_e^{1/2}
  E_{\mathrm{iso},53}^{1/2}\,\mathrm{days}, 
\end{equation}
which corresponds to 
\begin{equation}
  t_{t,\earth} = 3.4 \times 10^{4} (1+z) \epsilon_B^{2} \epsilon_e^{2}
  E_{\mathrm{iso},53} n_0\,\mathrm{days}.
\end{equation}

In the calculation present in this paper, we assume that $\epsilon_e =
0.1$, $\epsilon_B = 0.1$, and the GRB is located at $z = 1$.  At the
beginning of the simulation, $t_0 = 147$ days, the two break frequencies
are $\nu_{c0,\earth} = 2.8 \times 10^{12}\,\mathrm{Hz}$ and
$\nu_{m0,\earth} = 1.1 \times 10^{14}\,\mathrm{Hz}$.  Hence the GRB
outflow is initially in the fast cooling regime.  The transition from
the fast cooling to slow cooling regime takes place at 360 days in the
lab frame, which corresponds to 6.8 days for the observer, and
the transition frequency is $\nu_{t,\earth} = 4.6 \times
10^{11}\,\mathrm{Hz}$.

\begin{figure}
\plotone{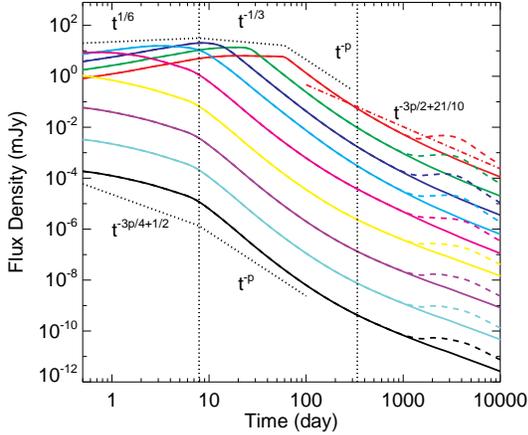}
\caption{Multi-frequency lightcurves.  The afterglow radiation from
  the forward jet is shown in solid lines, whereas the radiation from
  both the forward jet and counterjet is shown in dashed lines.  Flux
  density for various frequencies from radio to X-ray is plotted:
  $10^9$ ({\it red}), $10^{10}$ ({\it green}), $10^{11}$ ({\it blue}),
  $10^{12}$ ({\it cyan}), $10^{13}$ ({\it magenta}), $10^{14}$ ({\it
    yellow}), $10^{15}$ ({\it purple}), $10^{16}$ ({\it aqua}), and
  $10^{17}$ Hz ({\it black}).  Plotted in black dotted lines for
  comparison are lightcurves with slopes from \citet{Sari_PH_1999_ApJ,
    Rhoads_1999_ApJ}.  It should be noted that these analytic lines are
  arbitrary in magnitude.  An analytic lightcurve for frequency at 1 GHz
  in the nonrelativistic phase \citep{Frail_WK_2000_ApJ,
    Livio_Waxman_2000_ApJ} is also plotted in red dash-dotted line.
  The vertical dotted lines are at $7.9$ and $340$ days.
  \label{fig:lc}}
\end{figure}

\begin{figure}
\plotone{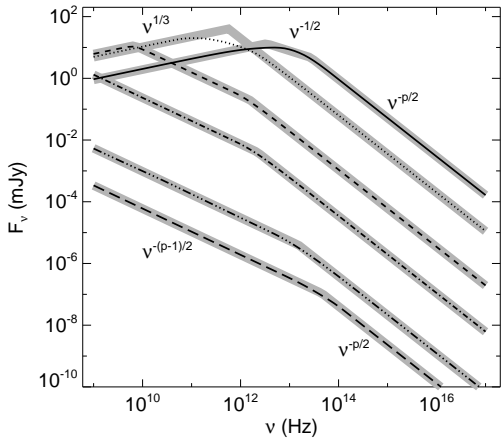}
\caption{Spectra at various times.  Thin black lines are results of
  our calculation for different times: $t_\earth = 0.6$ ({\it solid
    line}), $8$ ({\it dotted line}), $30$ ({\it dashed line}), $100$
  ({\it dash-dotted}), $1000$ ({\it dash dot dot line}), and $10000$
  days ({\it long dash line}). Thick gray lines are fitted
  using analytic formulae for synchrotron emission from nonthermal
  electrons with a power-law distribution. \citep{Sari_PN_1998_ApJ}.  
  \label{fig:sp}}
\end{figure}

\subsection{Results of Afterglow Calculation}
\label{sec:agres}

A large amount of data from the hydrodynamic simulation needs to be
stored for the calculation of afterglow radiation observed on the
earth.  Without an adequate number of data dumps, the afterglow
calculation will not be very accurate.  However, an unnecessary amount
of data would be generated if we simply dumped data equally spaced in
time because of the power-law time dependence of the blast wave.  For
our simulation, 3000 data dumps equally spaced in logarithmic time
have been made.  To test whether 3000 dumps is sufficient.  We have
performed low-resolution runs with 5000 dumps.  We found almost no
difference in the results of the afterglow calculation between using
all 5000 dumps and using only 3000 of the 5000 dumps.  Therefore we
conclude that the frequency of dumping data is sufficient.

In this calculation, we use a cosmology with $H_0 = 71
\mathrm{km}\,\mathrm{s}^{-1}\,\mathrm{Mpc}^{-1}$, $\Omega_M = 0.27$,
and $\Omega_\Lambda = 0.73$.  Using the hydrodynamic data
(\S~\ref{sec:hydro}) and the standard afterglow model
(\S~\ref{sec:agmodel}), we have calculated multi-frequency lightcurves
(Fig.~\ref{fig:lc}).  The calculated flux density covers a wide
frequency range: from $10^{9}\,\mathrm{Hz}$ (radio) to
$10^{17}\,\mathrm{Hz}$ (X-ray), and a wide range of observer time:
from 0.5 day after the burst to $\sim 27$ years.  The spectra at
various times are shown in Figure~\ref{fig:sp}.

\subsubsection{Jet Break}
\label{sec:jetbreak}

It is shown in Figure~\ref{fig:lc} that there is an achromatic break
in the lightcurves at $\sim 8$ days.  Using the analytic work of
\citet{Sari_PH_1999_ApJ}, we can estimate that the jet break happens
at $ t_j \approx 3.5 (1+z) E_{\mathrm{iso},53}^{1/3} n_0^{-1/3}
(\theta_0/0.2)^{8/3}\,\mathrm{days} \approx 7.0$ days.  However, we
believe that while they happened to get the correct answer, their
treatment of the sideways expansion was flawed (\S~\ref{sec:hydro}).
Furthermore, there is ambiguity in the relation between the emission
time and the observer time.  The jet break would still exist even if
there is no sideways expansion \citep{Meszaros_Rees_1999_MNRAS,
  Panaitescu_Meszaros_1999_ApJ, Granot_Kumar_2003_ApJ}.  Once the
observer sees the edge of the jet, the observed flux will decrease
rapidly due to the obvious difference between a spherical blast wave
and a jet and the sideways expansion will help decrease the flux.
It is usually thought that the break due to missing flux would be
shallower than the break due to sideways expansion.  This is incorrect
because the image of a GRB afterglow is limb-brighten
\citep{Granot_PS_1999_ApJ}.  We will discuss this in more detail
later in this section.  According to this line of argument, we can
also estimate the jet break time. For the sake of simplicity, one can
think that the information about the existence of an edge propagates
towards the axis as a rarefaction wave, which moves at the local sound
speed $c_s$.  Therefore, the angle of the head of the rarefaction wave
is $\theta_{\mathrm{RF}} \sim \theta_0 - \gamma^{-1} c_s/c \sim
\theta_0 - \gamma^{-1}/\sqrt{3}$.  An observer on the axis will see
the rarefaction when $\gamma \sim 1/\theta_{\mathrm{RF}} \sim
\theta_0^{-1} (1 + 1/\sqrt{3})$. The jet break time is around the
moment when that happens.  This leads to
\begin{equation}
  t_{j} \approx 3.9 (1+z) E_{\mathrm{iso},53}^{1/3} n_0^{-1/3}
  \left(\frac{\theta_0}{0.2}\right)^{8/3}\,\mathrm{days},
\end{equation}
where $E_{\mathrm{iso},53}$ is the isotropic equivalent energy in units of
$10^{53}\,\erg$.  

\begin{figure}
\plotone{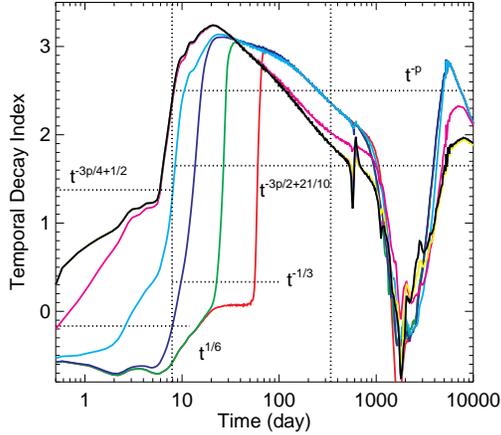}
\caption{Temporal decay index as a function of time.  Assuming $F_\nu
  \sim t^{-\alpha}$, the temporal decay index $\alpha$ for various
  frequencies from radio to X-ray is plotted: $10^9$ ({\it red}),
  $10^{10}$ ({\it green}), $10^{11}$ ({\it blue}), $10^{12}$ ({\it
    cyan}), $10^{13}$ ({\it magenta}), $10^{14}$ ({\it yellow}),
  $10^{15}$ ({\it purple}), $10^{16}$ ({\it aqua}), and $10^{17}$ Hz
  ({\it black}).  Note that the indices for the last four lines (from
  $10^{14}$ to $10^{17}$) are almost identical.  Also plotted for
  comparison are analytic results of \citet{Sari_PH_1999_ApJ,
    Frail_WK_2000_ApJ}.  The vertical dotted lines are at $7.9$ and
  $340$ days.
  \label{fig:alpha}}
\end{figure}

The analytic results of \citet{Sari_PN_1998_ApJ, Sari_PH_1999_ApJ,
  Rhoads_1999_ApJ} are also plotted in Figure~\ref{fig:lc} for
comparison.  For example, they predicted that the flux density at high
frequencies ($\nu_\earth > \nu_{m,\earth}$) evolves as
\begin{equation}
  F_\nu \propto t_\earth^{-p}. \label{eq:ftp}
\end{equation} 
Note that in our model the transition from the fast cooling to the
slow cooling regime takes place at $\sim 7$ days
(\S~\ref{sec:agmodel}; Fig.~\ref{fig:sp}), which is roughly the same
time as the jet break.  For $\nu_\earth < 10^{12}$ Hz, the flux
density evolves as $\sim t_\earth^{1/2}$ before the jet break,
deviating from the analytic result that it should evolve as $\sim
t_\earth^{1/6}$ \citep{Sari_PN_1998_ApJ} probably due to still
insufficient resolution for ultrarelativistic material, and then
becomes essentially flat until the frequency is above the typical
frequency $\nu_{m,\earth}$.  Then the flux density is slightly steeper
than $t_\earth^{-p}$.  For $\nu_{t,\earth} \sim 10^{12}$ Hz, the flux
density is essentially flat before the jet break, and is steeper than
$t_\earth^{-p}$ afterwards.  For $\nu_\earth > 10^{12}$, the flux
density is flatter than $t_\earth^{-3p/4+1/2}$ before the jet break,
probably also due to still insufficient resolution for
ultrarelativistic material ($\gamma > 10$). Then it becomes much
steeper than $t_\earth^{-p}$.

The most notable difference between our simulation results and
previous analytic results \citep{Sari_PH_1999_ApJ, Rhoads_1999_ApJ} is
the sharpness of the jet break in our results.  At radio frequencies,
the post-break lightcurves after the spectral break can be fitted
approximately by $\sim t_\earth^{-p}$.  But at optical and X-ray
frequencies, the post-break lightcurves can no longer be fitted by a
simple power-law.  If the power-law form $t^{-\alpha}$ is used for the
fit, the lightcurves have a varying temporal decay index as shown in
Figure~\ref{fig:alpha}.  The temporal indices for $\nu_\earth =
10^{14}\,\mathrm{Hz}$ and above are almost identical as expected for
frequencies above both the cooling frequency $\nu_{c,\earth}$ and
typical frequency $\nu_{m,\earth}$.  Note that a statistical study of
pre-Swift bursts \citep{Zeh_KK_2006_ApJ} found that the post-break
temporal decay index was in the range of $1.30 - 3.03$.  It is shown
in Figure~\ref{fig:alpha} that the post-break temporal decay index can
increase to $\sim 3$ and then decreases to below the electron
power-law index $p$.  This is consistent with the argument that jet
break is caused by the {\emph{missing}} flux outside the jet opening
angle.  It should be noted that photons received by an observer at a
given observer time are not emitted at the same time in the lab frame
because of the difference in light travel time for different parts of
the outflow.  It takes longer for photons emitted from off-axis parts
of the jet to reach the observer than photons emitted on the axis at
the same lab frame time.  Thus, at a given observer time, the
radiation from the off-axis part of the jet was emitted earlier when
the jet energy density (which is decreasing as the jet decelerates)
was higher and is brighter than that from the axis.  In other words,
the image of a GRB afterglow is limb-brightened
\citep{Granot_PS_1999_ApJ}.  Once the edge of the jet is seen by the
observer, the flux density will decrease rapidly.  Since the
{\emph{missing}} flux is brighter, the temporal decay index will
overshoot after the jet break \citep[see also][]{Granot_2007_RMxAC}.
Because the afterglow image is more limb-brightened for frequencies
above the cooling frequency than below \citep{Granot_PS_1999_ApJ}, the
overshooting is more prominent at higher frequencies than lower
frequencies.

The spectra at various times are shown in Figure~\ref{fig:sp}.  They
can be fit by broken power-law curves as expected.  This confirms the
analytic analysis in \S~\ref{sec:agmodel}.  The outflow is initially
in the fast cooling regime.  The two break frequencies become equal at
$\sim 8$ days.  Then the slow cooling regime is entered.  The
typical frequency $\nu_{m,\earth}$ continues to decrease over time.
However the cooling frequency $\nu_{c,\earth}$ increases slowly from
$\sim 10^{12}$ Hz at 8 days to $\sim 10^{14}$ Hz at 10,000 days.

\subsubsection{Nonrelativistic Regime}
\label{fig:nr}

\begin{figure}
\plotone{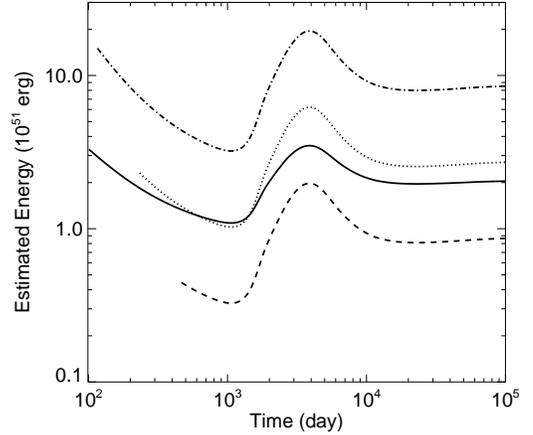}
\caption{Estimated energy as a function of observer time.  With
  observed flux density at $\nu_\earth = 1$ GHz, one can estimate the
  total energy of the GRB outflow using Eq.~\ref{eq:fnu_nr2} ({\it
    solid line}) or Eq.~\ref{eq:fnu_nr}.  The latter estimate requires
  an estimate of the Sedov-von Neumann-Taylor time $t_{\mathrm{SNT}}$.
  Three different values for $t_{\mathrm{SNT}}$ are used: 116 ({\it
    dotted line}), 232 ({\it dashed line}), and 58 days({\it
    dash-dotted line}).  The true energy is $ 2 \times 10^{51}\,\erg$.
  \label{fig:eradio}}
\end{figure}

We have discussed in \S~\ref{sec:hydro} that $t_{\mathrm{NR}}$ in the
lab frame marks the beginning of nonrelativistic regime.  Using the
relation $t_\earth \sim (1+z) t/4\gamma^2$, we can obtain
\begin{equation}
  t_{\mathrm{NR},\earth} \approx 170 (1+z)E_{\mathrm{iso},53}^{1/3} n_0^{-1/3}\,\mathrm{days}.
\end{equation}

When the GRB outflow becomes a Sedov-von Neumann-Taylor blast wave at
late times, the lightcurve will become flatter \citep{Dai_Lu_1999_ApJ,
  Frail_WK_2000_ApJ, Huang_Cheng_2003_MNRAS}.  At late times, the
radio frequency is typically in between $\nu_{m,\earth}$ and
$\nu_{c,\earth}$ (Fig.~\ref{fig:sp}).  The radio lightcurve in the
Sedov-von Neumann-Taylor phase evolves as \citep{Frail_WK_2000_ApJ,
  Livio_Waxman_2000_ApJ}
\begin{eqnarray}
  F_\nu &\approx& 0.2 (1+z)^{(3-p)/2}
  \left(\frac{\epsilon_e}{0.1}\right)
  \left(\frac{\epsilon_B}{0.1}\right)^{3/4} n_0^{3/4} E_{j,51}
  d_{L,28}^{-2} \nu_{\mathrm{GHz}}^{-(p-1)/2} \nonumber\\
  & & \times
  \left(\frac{t_\earth}{t_{\mathrm{SNT}}(1+z)}\right)^{(21-15p)/10}
  \,\mathrm{mJy}, \label{eq:fnu_nr}
\end{eqnarray}
here $d_{L,28}$ is the luminosity distance in units of $10^{28}\,\cm$,
and $\nu_{\mathrm{GHz}}$ is frequency in units of GHz.  

It should be emphasized that the GRB outflow at $t_{\mathrm{NR}}$
cannot be described by the Sedov-von Neumann-Taylor solution yet
because it is still highly nonspherical.  In fact, it does not become
approximately spherical until $\sim 5 t_{\mathrm{NR}}$.  Since the
transition from relativistic to nonrelativistic flow is very smooth
(\S~\ref{sec:hydro}), it is reasonable that the lightcurves at late
times become flat gradually (Figs.~\ref{fig:lc} and \ref{fig:alpha}).
There are no sharp breaks in them except the bumps from the counter
jet at very late times.  According to \citet{Frail_WK_2000_ApJ,
  Berger_KF_2004_ApJ, Frail_etal_2005_ApJ} since the lightcurve can be
observed months or even years after the burst when the outflow is
putatively a spherical Sedov-von Neumann-Taylor blast wave, the radio
afterglow would allow for an accurate estimate of the total energy of
the GRB outflow.  They used the flattening present in the observed
radio afterglow lightcurves to estimate $t_{\mathrm{SNT}}$, and
performed calorimetric analysis of the explosions.  However, there is
no sharp transition in the lightcurves calculated from our simulation
(Fig.~\ref{fig:lc}).  It would therefore be very inaccurate to
determine $t_{\mathrm{SNT}}$ from the rather smooth late-time
afterglow.  It is shown in Figure~\ref{fig:eradio} that the estimated
energy using Eq.~\ref{eq:fnu_nr} could have an order of magnitude
difference if the estimated $t_{\mathrm{SNT}}$ varies by a factor of
4.  Here, other parameters such as $\epsilon_e$, $\epsilon_B$, $p$ and
$n_0$ are assumed to be known.  The emission from the counter jet can
be observed at late times too.  This causes a bump in the lightcurves
(Fig.~\ref{fig:lc}).  Eq.~\ref{eq:fnu_nr} and our calculation do not
match very well until $ t_\earth \sim 50 (1+z) t_{\mathrm{SNT}} \sim
10^{4}$ days.  Fortunately, by substituting Eq.~\ref{eq:tsnt} into
Eq.~\ref{eq:fnu_nr}, we can obtain an equation without
$t_{\mathrm{SNT}}$,
\begin{eqnarray}
  F_\nu &\approx& 0.16 (1+z)^{(3-p)/2}
  \left(\frac{\epsilon_e}{0.1}\right)
  \left(\frac{\epsilon_B}{0.1}\right)^{3/4} n_0^{(29-10p)/20}
  E_{j,51}^{(5p+3)/10} \nonumber\\
  & & \times d_{L,28}^{-2} \nu_{\mathrm{GHz}}^{-(p-1)/2} 
  \left(\frac{t_{\earth,d}}{92 (1+z)}\right)^{(21-15p)/10}
  \,\mathrm{mJy}. \label{eq:fnu_nr2}
\end{eqnarray}
Here, a factor of $0.8$ is added to the expression to fit the
late-time afterglow better.  Figure~\ref{fig:eradio} shows that the
estimated energy using Eq.~\ref{eq:fnu_nr2} is more accurate.  Even at
$t \sim (1+z) t_{\mathrm{SNT}}/2 \sim 100 $ days, it only
overestimates the energy by $\sim 50\%$.  And the inferred energy
using Eq.~\ref{eq:fnu_nr2} agrees extremely well with the true energy
after $10^4$ days.

Fig.~\ref{fig:lc} shows that there is a bump in the lightcurves due to
radiation from the counter jet.  The late-time bump from the counter
jet has been discussed by \citet*{Granot_Loeb_2003_ApJ,
  Li_Song_2004_ApJ, Wang_HK_2009_arXiv}.  However, the bump in our
results is much smoother and broader than that in
\citet{Granot_Loeb_2003_ApJ, Li_Song_2004_ApJ}.  Our results are
consistent with those of \citet{Wang_HK_2009_arXiv}.  Initially, the
counter jet moves relativistically.  Then it slows down and becomes
nonrelativistic at $t_{\mathrm{NR}}$ in the lab frame.  Its radiation
cannot be observed by the observer on the earth until it becomes
nonrelativistic at $t_{\mathrm{NR}}$ in the lab frame.  This would
lead to an estimate of the time for the counter jet bump in the
lightcurve
\begin{equation}
  t_{\mathrm{cj},\earth} \approx 2(1+z) t_{\mathrm{NR}} \approx 1900
  (1+z) E_{\mathrm{iso},53}^{1/3} n_0^{-1/3}\,\mathrm{days},  \label{eq:tcj}
\end{equation}
and an estimate of the ratio of the two fluxes at the peak
\begin{equation}
  \frac{F_{\nu,\mathrm{cj}}}{F_{\nu,\mathrm{fj}}} \approx
  \left(\frac{1}{3}\right)^{(21-15p)/10} \approx 6,  
  \label{eq:ff}
\end{equation}
for $p = 2.5$.  In our calculation of the afterglow lightcurves, for
$\nu_\earth = 1$ GHz, the flux from the counter jet becomes comparable
to that from the forward jet at $\sim 1700$ days.  At $3800$ days, the
ratio of the two fluxes at 1 GHz reaches its peak of $6$.  The above
estimates (Eqs.~\ref{eq:tcj} and \ref{eq:ff}) agree extremely well
with the numerical results.  Because $t_{\mathrm{NR}}$ is the
nonrelativistic timescale in the lab frame for the fluid on the jet
axis, it is expected that the first appearance of the counter jet bump
from off-axis emission is earlier than $t_{\mathrm{cj},\earth}$.  And
numerical calculation indicates that the counter jet bump starts at
$\sim (1+z) t_{\mathrm{NR}}$.

\section{Discussion}
\label{sec:dis}

Our relativistic hydrodynamic simulation shows, as did
\citet{Granot_etal_2001_grba_conf}, that the sideways expansion of a
relativistic GRB jet is a very slow process (\S~\ref{sec:hydro}).
Analytic works \citep[e.g.,][]{Rhoads_1999_ApJ, Sari_PH_1999_ApJ}
based on a top-hat distribution of energy greatly overestimated the
rate of the sideways expansion.  A frequently asked question is, what
is the speed of the sideways expansion?  We think the question itself
is wrong because it implicitly assumes a top-hat distribution during
the expansion.  This would always lead to an exponential growth of the
jet opening angle.  The jet during the expansion is indeed far from a
top-hat distribution (Fig.~\ref{fig:dedo}), and it expands slowly.
Another question is, what causes the jet break?  Our calculations show
that the jet break in GRB afterglow lightcurves is mainly caused by
the {\emph{missing}} flux when the edge of the jet is observed
(\S~\ref{sec:jetbreak}).  Fortunately, the widely used formula, Eq.~1
in \citet{Sari_PH_1999_ApJ}, which relates the jet break time to the
properties of GRB jets is accurate.

It is generally believed that the spherical Sedov-von Neumann-Taylor
self-similar solution can be applied at $\sim t_{\mathrm{NR}}$
\citep[e.g.,][]{Piran_2005_RvMP}.  We, however, find that the time
$t_{\mathrm{NR}}$ is the beginning of a rather slow transition into
the Sedov-von Neumann-Taylor solution, and the outflow at that time is
still highly nonspherical (\S~\ref{sec:hydro}).  Our simulation shows
that the outflow can be described by the Sedov-von Neumann-Taylor
solution after $\sim 5 t_{\mathrm{NR}}$.  Note that the time
$t_{\mathrm{NR}}$ measured in the lab frame is not equivalent to the
observer time since the velocity of the flow is not sufficiently small
yet compared with the speed of light.  Fortunately again, the
afterglow at $t_\earth \sim (1+z) t_{\mathrm{NR}}$ is during the
nonrelativistic phase.

We have found that the lightcurve after the jet break will become
increasingly flatter over the time.  But the flattening is a very
gradual process.  There is no characteristic timescale at which the
lightcurve will suddenly become flatter.  Our results disagree with
the common notion that the flux density evolves as $\propto
t_\earth^{-p}$ and then switches to $\propto t_\earth^{(21-15p)/10}$
after $t_{\mathrm{NR}}$ or $t_s$.  

However, late-time radio observations can reveal a wealth of
information about the GRB outflow.  Eq.~\ref{eq:fnu_nr2} can be
potentially useful in determining the true energy of the outflow.  We
predict a late-time bump in flux density due to the radiation from the
counter jet part of the outflow \citep[see
also][]{Granot_Loeb_2003_ApJ, Li_Song_2004_ApJ, Wang_HK_2009_arXiv}.
The radio afterglow of GRB 030329 had been observed up to 1128 days
after the burst \citep{VanDerHorst_etal_2008_AA}.  No firm conclusion
was drawn as to whether or not the emission from the counter jet had
been observed.  However, the counter jet bump may have been observed
already in GRB 980703, which was monitored in the radio band up to $\sim
1000$ days \citep{Berger_KF_2001_ApJ}.  It is reasonable to attribute
the late-time radio flux to the host galaxy.  However, it is also
possible that the flux at $\sim 1000$ days was actually emitted by the
counter jet.  This possibility could be easily tested through a future
radio observation of the host galaxy of GRB 980703.  

At the end of the hydrodynamic simulation ($t \approx 150$ years), the
GRB outflow is almost a sphere with an aspect ratio of $\sim 0.8$.
Thus, it would be very difficult to distinguish such a GRB remnant at
an age of more than $\sim 200$ years from a supernova remnant using
morphology alone.  However, previous studies
\citep{Ayal_Piran_2001_ApJ,Ramirez-Ruiz_MacFadyen_2008_arXiv} have
found that a GRB remnant in a similar interstellar medium will not
become a sphere until $\sim 5000$ years.  The striking difference is
due to different initial setups.  In our hydrodynamic simulation, the
GRB outflow initially moves along the radial direction of spherical
coordinates, whereas it moves parallel to the symmetric axis in their
simulations.  Obviously, the outflow sweeps up the surrounding medium
at a much higher rate in our simulation than in theirs.  Therefore, it
is not surprising that the GRB outflow becomes a sphere at much
earlier time in our simulation than in theirs.

High-resolution hydrodynamic simulations of GRB outflows are very
computationally expensive.  An intermediate approach to model the
hydrodynamic evolution of GRB outflows could be as follows.  A
nonspreading jet obeying the Blandford-McKee solution can be assumed
until $t_{\mathrm{NR}}$.  For $t>5t_{\mathrm{NR}}$, the outflow can be
assumed to obey the Sedov-von Neumann-Taylor solution.  Interpolation
can be used for the transition phase $t_{\mathrm{NR}} < t <
5t_{\mathrm{NR}}$ (see Fig.~\ref{fig:slice}).

Many aspects of our numerical calculations can be improved.  The
calculation of radiation could include synchrotron self-absorption,
which can be important for radio afterglow, and the inverse Compton
process, which can be important for X-ray afterglow
\citep{Sari_Esin_2001_ApJ, Harrison_etal_2001_ApJ}.  In our numerical
simulation, the hydrodynamic evolution of the GRB outflow is
adiabatic.  This is a good approximation provided that $\epsilon_e \ll
1$ during the early fast cooling regime.  However, radiative loss
could greatly affect the hydrodynamics of the outflow during the early
afterglow phase (e.g., less than a day) if $\epsilon_e$ is close to 1
\citep*{Cohen_PS_1998_ApJ}.

A uniform jet described by the Blandford-McKee solution is assumed as
the initial condition of our numerical simulation.  This is justified
because the jet initially moves at ultrarelativistic speeds and
sideways expansion at very early times is likely to be modest.
Besides uniform jet models based on the Blandford-McKee solution,
structured jet models \citep*{Meszaros_RW_1998_ApJ, Perna_SF_2003_ApJ,
  Granot_Kumar_2003_ApJ} have also been proposed to explain various
phenomena, including jet break \citep*{Dai_Gou_2001_ApJ,
  Rossi_LR_2002_MNRAS, Zhang_Meszaros_2002_ApJ}.  Moreover, numerical
simulations have shown that nonuniform structures are expected for
relativistic jets emerging from massive stars
\citep*{Zhang_WM_2003_ApJ, Zhang_WH_2004_ApJ, Morsony_LB_2007_ApJ}.
Recently \citet{Gruzinov_2007} has shown that the spherical
Blandford-McKee solution is not an attractor for a generic asymmetric
explosion.  Therefore, it is very important to investigate
non-Blandford-McKee models and their consequences on GRB afterglows
with high-resolution numerical simulations, which we will present in
future publications.  We speculate that the shallow decay phase in the
early X-ray afterglows of Swift bursts
\citep[e.g.,][]{Nousek_etal_2006_ApJ,OBrien_etal_2006_ApJ} could be
due to non-Blandford-McKee behavior of the outflow and the normal
decay phase is reached as the outflow approaches the Blandford-McKee
solution.

Magnetic field is not included in the two-dimensional relativistic
hydrodynamic simulation presented in this paper, because we are mainly
concerned with the stage when the magnetic field is no longer
dynamically important.  However, since GRB jets may be powered by
electromagnetic processes \citep[e.g.,][]{Blandford_Znajek_1977_MNRAS,
  Uzdensky_MacFadyen_2006_ApJ, Uzdensky_MacFadyen_2007_ApJ,
  Barkov_Komissarov_2008_MNRAS, Bucciantini_etal_2009_arXiv}, magnetic
field might have important effects on the dynamics of the jet during
its pre-Blandford-McKee phase and early afterglows
\citep[e.g.,][]{Zhang_Kobayashi_2005_ApJ}.  Preliminary
one-dimensional relativistic magnetohydrodynamic simulations have
found that the early afterglows are strong dependent on the
magnetization of the GRB outflow \citep{Mimica_GA_2009_AA}.

The environment of a GRB can have a huge impact on its afterglow
\citep[e.g.,][]{Meszaros_RW_1998_ApJ, Chevalier_Li_1999_ApJ,
  Chevalier_Li_2000_ApJ}.  Thus, another important issue is to
investigate GRB outflows propagating in stellar winds since the
progenitors of long soft GRBs are believed to be massive stars
\citep{Woosley_1993_ApJ, Paczynski_1998_ApJ,
  MacFadyen_Woosley_1999_ApJ}.  More recently, late-time Chandra
observations of the X-ray afterglow of GRB 060729 up to 642 days after
the burst have been reported by \citet{Grupe_etal_2009_arXiv}.  It is
interesting that the lightcurve at such late times shows signs of
steepening possibly due to either jet break or spectral break rather
than flattening due to nonrelativistic motion.  This might indicate
that the onset of the Sedov-von Neumann-Taylor phase has been decayed
due to the wind medium environment

We adopted the standard approach in modeling GRB afterglow radiation.
For example, the magnetic field is assumed to carry constant fractions
of the internal energy density, and so do synchrotron emitting
nonthermal relativistic electrons.  Moreover, the fractions are
assumed to be constant throughout the entire afterglow phase.  The
parameterization is a result of the lack of understanding of these
processes.  It remains unclear how magnetic fields are generated and
how particles are accelerated in GRB outflows.  The Weibel instability
is a plausible mechanism \citep{Gruzinov_Waxman_1999_ApJ,
  Medvedev_Loeb_1999_ApJ}.  Recent plasma simulations of the Weibel
instability in relativistic collisionless shocks have shown promising
results \citep{Nishikawa_etal_2003_ApJ, Spitkovsky_2008_ApJ}.  More
recently, \citet*{Zhang_MW_2009_ApJL} have demonstrated amplification
of magnetic field by a turbulent dynamo triggered by the
Kelvin-Helmholtz instability with three-dimensional relativistic
magnetohydrodynamic simulations.  For conditions relevant for late
afterglow and prompt GRB emission \citet{Zhang_MW_2009_ApJL} obtained
$\epsilon_B \sim 5\times10^{-3}$.

In this paper, we have calculated afterglows for an observer on the
axis of the GRB jet. However, orphan afterglows are expected for an
off-axis observer, who has missed the main burst because of the
relativistic beaming effect \citep{Rhoads_1997_ApJ}.  The rate of
orphan afterglows have been estimated analytically
\citep{Dalal_GP_2002_ApJ, Granot_etal_2002_ApJ,
  Levinson_etal_2002_ApJ, Nakar_PG_2002_ApJ}.  Thus far, no orphan
afterglow has been detected.  A recent survey of 68 local Type Ibc
supernovae, including 6 broad-lines supernovae also known as
``hypernovae'' found no evidence of off-axis GRBs
\citep{Soderberg_etal_2006_ApJ}.  Note that we have found that the
sideways expansion has been overestimated in previous analytic works.
Thus, previous analytic works tend to overestimate the rate of orphan
afterglows.  The issue of off-axis GRB afterglows will be investigated
in a future publication.

\acknowledgments

We would like to thank Joseph Gelfand, Jonathan Granot, Andrei
Gruzinov, Enrico Ramirez-Ruiz and Bing Zhang for many stimulating
discussions.  We appreciate helpful comments on the manuscript from
Jonathan Granot and Bing Zhang.  The RAM code used in this work was in
part based upon the FLASH code developed by the DOE-supported
ASCI/Alliance Center for Astrophysical Thermonuclear Flashes at the
University of Chicago.  Specifically, we used the PARAMESH AMR and I/O
tools from FLASH version 2.3.

\bibliography{afterglow}

\end{document}